\documentclass[structabstract]{aa}
\usepackage{txfonts}
\usepackage[utf8]{inputenc}
\usepackage[T1]{fontenc}
\usepackage{natbib}
\usepackage[breaklinks=true]{hyperref}
\usepackage{amssymb}
\usepackage{graphicx}
\usepackage{booktabs} 

\bibpunct{(}{)}{;}{a}{}{,} 

\title{The properties of non-thermal X-ray filaments in young supernova remnants}
\author{R. Rettig\inst{1} \and M. Pohl\inst{1,2}}
\institute{Institute of Physics and Astronomy, University of Potsdam, 14476 Potsdam, Germany \and DESY, 15738 Zeuthen, Germany}
\date{Received 13 April 2012 / Accepted 26 July 2012}
\abstract{Young supernova remnants (SNRs) exhibit narrow filaments of non-thermal X-ray emission whose widths can be limited either by electron energy losses or damping of the magnetic field.}{We want to investigate whether or not different models of these filaments can be observationally tested.}{Using observational parameters of four historical remnants, we calculate the filament profiles and compare the spectra of the filaments with those of the total non-thermal emission. For that purpose, we solve an one-dimensional stationary transport equation for the isotropic differential number density of the electrons.}{We find that the difference between the spectra of filament and total non-thermal emission above 1~keV is more pronounced in the damping model than in the energy-loss model.}{A considerable damping of the magnetic field can result in an observable difference between the spectra of filament and total non-thermal emission, thus potentially permitting an observational discrimination between the energy-loss model and the damping model of the X-ray filaments.}

\begin{document}

\titlerunning{Non-thermal X-ray filaments in young supernova remnants}
\authorrunning{R. Rettig \& M. Pohl}
\maketitle

\section{Introduction}

Based on simple energetic considerations regarding the energy density of cosmic rays and the energy release per supernova explosion, SNRs have long been thought to be sources of galactic cosmic rays. This presumption is supported by numerous detections of non-thermal emission in the radio and X-ray band \citep[e.g.,][]{1995Natur.378..255K,1999ApJ...525..357S,2001ApJ...548..814S,2000PASJ...52.1157B} observed from the direction of known SNRs and interpreted to be synchrotron radiation of relativistic electrons with energies up to $100\,$TeV.

High-resolution observations of young SNRs performed with the \emph{Chandra} satellite show that this emission of non-thermal radiation is concentrated in narrow regions on the limbs \citep{2003ApJ...584..758V,2005ApJ...621..793B}. These regions of increased synchrotron emissivity close to the forward shock are called filaments and demonstrate the presence of high-energy electrons around their acceleration sites.

The most plausible process for the acceleration of electrons is diffusive shock acceleration (DSA), which leads to a power-law distribution of particles \citep[e.g.,][]{1978MNRAS.182..147B,1978ApJ...221L..29B,1983RPPh...46..973D}. Although no clear evidence for relativistic-ion acceleration exists at shocks, the DSA-mechanism is also considered to be responsible for the acceleration of cosmic-ray nuclei, as indicated by observations of non-relativistic ion acceleration at solar-wind shocks driven by coronal mass ejections \citep{2011ApJ...735....7R}. However, many details of the DSA are still vague such as the maximum energy of particles, the role of the magnetic field, and how the particles are injected into the acceleration process (also referred to as the injection problem). Apparently, the investigation of the properties of the non-thermal filaments may provide key information for a better understanding of the DSA-mechanism. In particular, knowing the magnetic-field strength gives constraints on the maximum particle energy achievable in the acceleration process, which can help answering the question whether SNRs can accelerate particles to energies above the knee in the cosmic-ray spectrum.   

Accurate analysis of several SNRs shows that the filamentary structures are very thin compared with the radii of the remnants. This limitation of the filament widths is associated with a rapid decrease of the synchrotron emissivity that can be explained by energy losses of the electrons in a locally enhanced magnetic field. A number of authors have used that model to constrain the magnetic-field strength, the degree of turbulence and the obliquity \citep[e.g.,][]{2003ApJ...589..827B,2006A&A...453..387P,2010ApJ...714..396A}. \citet{2010ApJ...714..396A}, for instance, have investigated several filaments of the remnant of Cas A and found that the magnetic fields of the filaments are highly turbulent and nearly perpendicular to the shock normal. Another important result of this and other studies is an estimate of the downstream magnetic-field strength that is higher than simple shock compression would suggest. Such observations indicate an additional amplification of the magnetic field in the shock region of the SNRs. A possible amplification process could be a streaming instability in the upstream region as proposed by \citet{2000MNRAS.314...65L} and \citet{2004MNRAS.353..550B}, or the effects of preexisting turbulent density fluctuations on the propagating shock front \citep{2007ApJ...663L..41G}. 

Besides energy losses, also the magnetic field itself can limit the filament widths. Based on the turbulence relaxation downstream of the forward shock and neglecting any amplification process, \citet{2005ApJ...626L.101P} have calculated that the turbulent magnetic field downstream can decay exponentially on a damping-length scale $l_{\mbox{\tiny d}}=10^{16}-10^{17}\,$cm that is small enough to produce the narrow observable filaments. Furthermore, from observations of the post-shock steepening of the synchrotron spectrum in Tycho's SNR it can be seen that also the damping of the magnetic field fairly well describes the corresponding X-ray data \citep{2007ApJ...665..315C}, and thus, may appear within the filaments.

Since the magnetic field controls the radiative cooling of electrons, high magnetic fields lead to strong cooling, and too few high-energy electrons remain capable of producing the gamma-ray emission, that is observed from regions near the edges of numerous SNRs \citep[e.g.,][]{2007ApJ...661..236A,2010A&A...516A..62A,2011ApJ...734...28A}. Any gamma rays observed in such a case are likely to be hadronic in origin. Weak cooling leads to a large number of high-energy electrons and the possibility of gamma-ray emission through inverse Compton or bremsstrahlung processes. All these implications of the magnetic-field structure on the particle acceleration, gamma-ray emission and magnetic-field amplification make it necessary to understand the non-thermal filaments in detail.

In this paper we investigate the properties of the filaments for both cases, filaments limited by electron energy losses or by damping of the magnetic field. For that purpose, using observational values of some characteristical SNR parameter, we calculate the X-ray emission of the filaments. The resulting filament profiles then allow us to make specific predictions regarding the magnetic-field strength. We additionally calculate the total non-thermal emission, which shall be referred to as "plateau", and compare their spectra with those of the filaments. It should be noted that in our models we only consider non-thermal synchrotron emission and restrict ourselves to the evolution of relativistic electrons in the downstream region. Furthermore, we assume the electrons to be already accelerated at the shock front and treat our problem to be independent of the acceleration process. We also do not consider any electron propagation into the upstream region and simplify the SNRs to be spherical objects of constant downstream-velocity profile. Recent hydrodynamical simulations suggest that this oversimplification of a constant velocity is an acceptable approximation only for SNRs of an age of less than several hundred years \citep{2012APh....35..300T}, implying that our models are restricted to SNRs being in the adiabatic phase and just entering the Sedov phase, respectively. 

\section{Modelling the filaments}

We calculate the X-ray intensity as a function of the projected radius, $r_{\mbox{\tiny p}}$. It is an integral over the synchrotron emission coefficient, $j_\nu$, along the line of sight, $I_\nu=\int_{-\infty}^{\infty} j_\nu \, \mbox{d}y$. Using $r^2=y^2+r_{\mbox{\tiny p}}^2$ and taking into account that only emission originating inside the SNR contributes, the X-ray intensity can be written as
\begin{equation} \label{eqn: intensity distribution}
I_{\nu}(r_{\mbox{\tiny p}})=2\int_{r_{\mbox{\tiny p}}}^{r_{\mbox{\tiny s}}} \, \frac{j_{\nu}(r)}{\sqrt{1-\frac{r_{\mbox{\tiny p}}^2}{r^2}}} \, \mbox{d}r \; ,
\end{equation}
where $r$ and $r_{\textrm{\tiny s}}$ denote the positions inside the SNR and the radius of the SNR, respectively. Then, obtaining the corresponding spectrum of the filaments just involves an integration over the X-ray emission along the projected radius, whereas the spectrum of the plateau emission can be calculated as a volume integral over the synchrotron emission coefficient. In both cases, the isotropic synchrotron emission coefficient is needed, given by
\begin{equation} 
j_{\nu}(r)=\frac{1}{4\pi}\int_{}^{\infty} N(r,E)P_{\nu}(r,E)\, \mbox{d}E \; ,
\end{equation}
where $N(r,E)$ and $P_{\nu}(r,E)$ are the isotropic differential electron number density and the spectral emissivity per electron, respectively. Thus, we need to derive the appropriate electron distribution within the filaments.

\subsection{The electron distribution}

In the following, we derive the electron distribution that is necessary to calculate the synchrotron emissivity. To do this, a transport equation describing the dynamics of a distribution of relativistic electrons affected by advection, diffusion and energy losses needs to be solved. According to simulations described in \citet{2012APh....35..300T}, we can approximate the advection velocity of young SNRs to be constant downstream of the shock, implying that energy losses due to adiabatic deceleration can be neglected. Note that the non-thermal emission come from a thin spherical shell near the edges of the SNRs. If we restrict our treatment to a region near the shock that is crossed by the advection flow on a timescale short compared with the age of the SNR, then we approximate the electron distribution with a one-dimensional steady-state solution.

It is convenient to introduce a comoving spatial coordinate, $z=r_{\mbox{\tiny s}}-r$, where $z=0$ marks the position of the shock front at all times. Hence, restricting ourselves on the downstream region, the one-dimensional transport equation for the isotropic differential number density, $N=N(z,E)$, reads as follows:
\begin{equation} \label{eqn:transport equation}
\frac{\partial}{\partial z}\left[D(z,E)\frac{\partial N}{\partial z}\right]-v\frac{\partial N}{\partial z}-\frac{\partial}{\partial E}\left[\beta(z,E)N\right]+Q(z,E)=0 \,.
\end{equation}
In this equation $v$ denotes the constant advection velocity of the electrons relative to the forward shock, $D(z,E)$ is the diffusion coefficient, $\beta(z,E)=\mbox{d}E/\mbox{d}t$ represents the electron energy loss due to the emission of radiation, and $Q(z,E)$ is the source term describing the injection of accelerated electrons into the propagation process.

Since the electrons are likely accelerated by the DSA-mechanism at the forward shock ($z=0$), we assume the electrons to be injected with a power-law dependence $E^{-s}$, where $s$ is the injection index. Hence, the source term reads
\begin{equation} \label{eqn:source term}
Q(z,E)=q_0E^{-s}\exp\left(-\frac{E}{E_{\mbox{\tiny cut}}}\right)\delta(z) \;,
\end{equation}
and includes a cut-off at energy $E_{\mbox{\tiny cut}}$, because the maximum possible energy can be limited by either the finite acceleration time of the SNR \citep{1991MNRAS.251..340D} or energy losses \citep{1984A&A...137..185W}.

Eq. (\ref{eqn:transport equation}) can be solved using Green's method, implying that the solution can be written in terms of Green's function,
\begin{equation} \label{eqn:solution}
N(z,E)=\int_0^{\infty}\mbox{d}z'\int_{}^{\infty}\mbox{d}E' \, g(z,z';E,E')Q(z',E') \;,
\end{equation}
where Green's function $g=g(z,z';E,E')$ satisfies
\begin{equation} \label{eqn:transport equation with Green's function}
\frac{\partial}{\partial z}\left[D(z,E)\frac{\partial g}{\partial z}\right]-v\frac{\partial g}{\partial z}-\frac{\partial}{\partial E}\left[\beta(z,E)g\right]=-\delta(z-z')\delta(E-E') \;.
\end{equation}

Assuming the diffusion coefficient and the energy losses to be separable in a spatial and in an energetic part,
\begin{equation}
D(z,E)=d(z)D(E) \; ,
\end{equation}
\begin{equation}
\beta(z,E)=-b(z)B(E) \; ,
\end{equation}
and that the spatial dependent terms are inversely proportional to each other,
\begin{equation} \label{eqn:alpha}
d(z)b(z)=\alpha=\mbox{const.} \; ,
\end{equation}
as well as introducing the substitutions
\begin{equation} \label{eqn: substitution Green's function}
g(z,z';E,E')=\frac{G(z,z';E,E')}{B(E)} \; ,
\end{equation}
\begin{equation} \label{eqn: substitution x}
x(z)=\int_0^z \frac{\mbox{d}y}{d(y)} \; ,
\end{equation}
\begin{equation} \label{eqn:substitution lamba}
\lambda(E)=\frac{1}{\alpha}\int_{E}^{\infty} \frac{\mbox{d}u}{B(u)} \; ,
\end{equation}
an analytical solution to Eq. (\ref{eqn:transport equation with Green's function}) can be found in the literature \citep[][see their Eq. (A20)]{1980ApJ...239.1089L} and reads 
\begin{eqnarray} \label{eqn:Green's function}
G(x,x';\lambda,\lambda')&=&\frac{\Theta(\lambda-\lambda')}{2\alpha\sqrt{\pi}}\sqrt{\frac{1}{\int_{\lambda'}^{\lambda}\, D(t)\,\mbox{d}t}} \nonumber \\
&& \times \exp\left\{-\frac{\left[v(\lambda-\lambda')+x'-x\right]^2}{4\int_{\lambda'}^{\lambda}\, D(t)\, \mbox{d}t}\right\} \; ,
\end{eqnarray}
where $\Theta(\lambda-\lambda')$ is the step function and $\lambda'=\lambda(E')$. Note, this analytical solution is valid only if Eq. (\ref{eqn:alpha}) applies.

In general, the diffusion coefficient is an unknown parameter. But it is often assumed that the diffusion proceeds in the so-called Bohm-regime. Hence, the diffusion coefficient can be written in the extreme-relativistic limit $E\gg mc^2$ as $D=\eta r_{\mbox{\tiny L}}c/3$, where $r_{\mbox{\tiny L}}=E/(eB)$ and $\eta\geq 1$ are the gyroradius and gyrofactor, respectively. Here, $m$ is the mass of the electron, $e$ is its charge, $c$ is the speed of light, and $B$ is the magnetic-field strength. Therefore, we may write $D(E)=D_0E$, where $D_0=\eta c/(3eB)$. Additionally, we assume the emission of synchrotron radiation to be the main energy-loss process that is proportional to the square of the electron energy, $B(E)=b_0E^2$, where $b_0=4e^4B^2/(9m^4c^7)$. Using these assumptions, Eq. (\ref{eqn:substitution lamba}) rewrites as $$\lambda(E)=\frac{1}{\alpha}\int_E^{\infty}\frac{\mbox{d}u}{b_0u^2}=\frac{1}{\alpha b_0 E}\, ,$$ so that $$D(E)=\frac{D_0}{\alpha b_0 \lambda}$$ and hence, $$\int_{\lambda'}^{\lambda}\, D(t) \, \mbox{d}t=\frac{D_0}{\alpha b_0}\ln\left(\frac{\lambda}{\lambda'}\right)=\frac{D_0}{\alpha b_0}\ln\left(\frac{E'}{E}\right) \; .$$ According to Eq. (\ref{eqn: substitution Green's function}) and Eq. (\ref{eqn:Green's function}), Green's function then reduces to
\begin{eqnarray}
g(x,x';E,E')&=&\frac{\theta(E'-E)}{2\sqrt{\pi\alpha b_0 D_{0}}E^{2}}\sqrt{\frac{1}{\ln\left(\frac{E'}{E}\right)}} \nonumber \\
&& \times \exp\left\{ -\frac{\left[\frac{v}{\alpha b_0}\left(\frac{1}{E}-\frac{1}{E'}\right)+x'-x\right]^{2}}{\frac{4D_{0}}{\alpha b_0}\ln\left(\frac{E'}{E}\right)}\right\} \, .
\end{eqnarray} 

Now, the electron distribution can be easily determined using Eq. (\ref{eqn:solution}) and the appropriate source distribution (\ref{eqn:source term}), so that
\begin{eqnarray}
 N(x(z),E) &=& \int_{x(0)}^{^{x(\infty)}}\mbox{d}x'\int_{}^{\infty} \mbox{d}E' \, \frac{q_{0}\theta(E'-E)}{2\sqrt{\pi\alpha b_0 D_{0}}}\,E^{-2}E'^{-s}\sqrt{\frac{1}{\ln\left(\frac{E'}{E}\right)}} \nonumber \\
 && \times \exp\left\{ -\frac{E'}{E_{\mbox{\tiny cut}}}-\frac{\left[\frac{v}{\alpha b_0}\left(\frac{1}{E}-\frac{1}{E'}\right)+x'-x\right]^{2}}{\frac{4D_{0}}{\alpha b_0}\ln\left(\frac{E'}{E}\right)}\right\} \delta(x') \nonumber \\
 &=& \frac{q_{0}}{2\sqrt{\pi\alpha b_0 D_{0}}}\,E^{-2}\int_{E}^{\infty}\frac{E'^{-s}}{\sqrt{\ln\left(\frac{E'}{E}\right)}} \nonumber \\
 & &  \times \exp\left\{-\frac{E'}{E_{\mbox{\tiny cut}}}-\frac{\left[\frac{v}{\alpha b_0 E}\left(1-\frac{E}{E'}\right)-x(z)\right]^{2}}{\frac{4D_{0}}{\alpha b_0}\ln\left(\frac{E'}{E}\right)}\right\} \mbox{d}E' \; .
\end{eqnarray}
Substituting $n=E'/E$, we can rewrite the desired result as
\begin{eqnarray} 
N(x(z),E) &=& \frac{q_{0}}{2\sqrt{\pi\alpha b_0 D_{0}}}\,E^{-(s+1)}\int_{1}^{\infty}\frac{n^{-s}}{\sqrt{\ln(n)}} \nonumber \\
               & & \times \exp\left\{-\frac{nE}{E_{\mbox{\tiny cut}}}-\frac{\left[\frac{v}{\alpha b_0 E}\left(1-\frac{1}{n}\right)-x(z)\right]^{2}}{\frac{4D_{0}}{\alpha b_0}\ln(n)}\right\} \mbox{d}n \; . \label{eqn:electron distribution}
\end{eqnarray}   

\subsection{Parameter used in the models}

The models we use are based on different assumptions on the magnetic field. In the first model, which shall be referred to as "energy-loss model", we assume the magnetic field to be spatially constant, whereas the second one, which is referred to as "magnetic-field damping model", includes the damping of magnetic turbulence and assumes a spatially-dependent magnetic-field strength described by a profile following the relation 
\begin{equation} \label{eqn:magnetic field}
B(z)=B_{\mbox{\tiny min}}+(B_{\mbox{\tiny max}}-B_{\mbox{\tiny min}}){\mbox{}}\exp\left(-\frac{z}{l_{\mbox{\tiny d}}}\right) \; ,
\end{equation}
where $l_{\mbox{\tiny d}}$ is the damping length and $z\geq 0$. Here, we choose the minimum value of the magnetic field to be similar to that of the interstellar medium, $B_{\mbox{\tiny min}}=10~\mu$G, whereas the maximum value, $B_{\mbox{\tiny max}}$, corresponds to the field at the shock. It should be noted that Eq. (\ref{eqn:magnetic field}) describes the averaged amplitude of the magnetic field in a given volume. Hence, we do not make any assumption on the magnetic-field direction and do not distinguish between parallel and transverse diffusion, as is done in detailed calculations of diffusion coefficients \citep[e.g.,][]{2006A&A...453..193M}.

According to the Rankine-Hugoniot conditions, the downstream advection velocity can be expressed by the shock velocity, $v_{\mbox{\tiny s}}$, through $v=v_{\mbox{\tiny s}}/4$, if we consider strong shocks with a high Mach number and a monatomic gas with adiabatic index of 5/3, leading to a shock compression ratio of 4. Thus, we neglect any non-linear effects expected to occur with efficient particle acceleration, which would modify the shock \citep{2004A&A...413..189E}. Furthermore, the filament width, $w$, is defined as the length, at which the intensity described by Eq. (\ref{eqn: intensity distribution}) is reduced by a factor $1/e$ of its maximum.

To have a realistic model, the values chosen for the shock velocity, filament width, and radius are based on reference values of real SNRs. In our case, we consider the young remnants of the historical supernovae SN 1006, Cas A, Tycho and Kepler. It should be noted that the real filament widths found in the literature have been measured in a certain X-ray band and not for an individual X-ray energy. However, as is shown in the calculation done in Sect. \ref{sect:results}, the shape of the filament profiles depends on the X-ray energy. Nevertheless, we relate the observational value of the width to a X-ray energy of 5 keV, since it is an energy, at which no significant contribution from thermal emission is expected.

Additionally, we treat the cut-off energy to be the maximum electron energy that can be achieved in the acceleration process. Unlike cosmic-ray nuclei whose energy is probably age-limited due to the finite acceleration time available, we assume the maximum electron energy to be loss-limited, since the electrons experience synchrotron losses during their acceleration. By equating the acceleration time scale to the synchrotron loss time, it is possible to derive an expression for the maximum electron energy in terms of the downstream magnetic-field strength and shock velocity \citep{2006A&A...453..387P}:
\begin{equation}\label{eqn:cutoff energy}
E_{\mbox{\tiny cut}}\equiv E_{\mbox{\tiny max}}\simeq (8.3~\mbox{TeV})\,\eta^{-1/2}\left(\frac{B(z=0)}{100~\mu\mbox{G}}\right)^{-1/2}\left(\frac{v_{\mbox{\tiny s}}}{1000~\mbox{km/s}}\right)\;.
\end{equation}
Since different mechanism can account for magnetic-field amplification \citep{2000MNRAS.314...65L, 2007ApJ...663L..41G}, the structure of the magnetic field is generally unknown within the shock region. We therefore simplify the problem by making the assumption that only the magnetic-field strength at the shock determines the maximum electron energy given by Eq. (\ref{eqn:cutoff energy}). In addition, we have taken a shock compression ratio of 4.

At last, we assume the injection index to be $s=2$, which results from the Rankine-Hugoniot conditions for strong shocks \citep{1978MNRAS.182..147B}, as well as Bohm diffusion ($\eta=1$), which implies the smallest possible value of the diffusion coefficient, as the mean free path of the particle is equal to the gyroradius.

All parameter used are summarized in Table \ref{tab:parameters}. Note, that it is possible that all four SNRs could exhibit similar shock velocities. To take the uncertainties of this quantity into account, we perform the calculation of SN 1006 and Kepler for two different shock velocities.

\begin{table*} 
\caption{Parameters used to derive the filament profiles and spectra. Furthermore, $s=2$ and Bohm diffusion are assumed in all SNRs.} \label{tab:parameters}
\centering
\begin{tabular}{c c c c c c c c c c c c}
\hline \hline
SNR & Age & \multicolumn{2}{c}{Distance} & \multicolumn{2}{c}{$r_{\mbox{\tiny s}}$} & \multicolumn{2}{c}{$v_{\mbox{\tiny s}}$} & \multicolumn{3}{c}{$w$} & $B_{\mbox{\tiny min}}$ \\ \cmidrule(lr){3-4} \cmidrule(lr){5-6} \cmidrule(lr){7-8} \cmidrule(lr){9-11}
    & [yr] & [kpc] & Reference & [arcmin]\tablefootmark{a} & [pc]\tablefootmark{b} & [km\,$\mbox{s}^{-1}$] & Reference & [arcsec] & [pc]\tablefootmark{b} & Reference & [$\mu$G]\tablefootmark{c} \\
\hline
SN 1006\tablefootmark{d} & 1000                  & 2.2   & 1  & 15  & 10  & 4900   & 2  & 20  & 0.2    & 3    & 10 \\
                         &                       &       &    &     &     & (2900) &    &     &        &      &    \\
Cas A                    &  330\tablefootmark{e} & 3.4   & 4  & 2.5 & 2.5 & 5200   & 5  & 1.5 & 0.03   & 6, 7 & 10 \\
Tycho                    &  440                  & 2.5   & 8  & 4   & 3   & 5000   & 9  & 5   & 0.06   & 6    & 10 \\
Kepler\tablefootmark{f}  &  410                  & 4.8   & 10 & 1.5 & 2   & 5040   & 11 & 3.5 & 0.08   & 6    & 10 \\                                                                      
                         &                       & (6.4) &    &     & (3) & (6720) &    &     & (0.11) &      & \\
\hline
\end{tabular}
\tablefoot{
\tablefoottext{a}{Taken from \citet{2009BASI...37...45G}.}
\tablefoottext{b}{Directly inferred using the distance and the appropriate quantity given in angular units.} 
\tablefoottext{c}{Only used in the magnetic-field damping model.}
\tablefoottext{d}{The shock velocity in the northwestern limb (value in brackets) is used, as well as in the northeastern limb, where strong electron acceleration appears to occur.}
\tablefoottext{e}{The supernova explosion may have been observed in 1680 \citep{1980JHA....11....1A}. Otherwise, the detection of radioactive $^{44}$Ti \citep{1997NuPhA.621...83H} and the analysis of the dynamics of this remnant \citep{2006ApJ...645..283F} also suggest an explosion date in the late 17th century.}
\tablefoottext{f}{Distance to this remnant is uncertain. Calculation is performed for the lower and upper limit (values in brackets) to the distance.}
}
\tablebib{
(1)~\citet{2003ApJ...585..324W}; (2)~\citet{2002ApJ...572..888G} + \citet{2009ApJ...692L.105K}; (3)~\citet{2003ApJ...589..827B}; (4)~\citet{1995ApJ...440..706R}; (5)~\citet{1998A&A...339..201V}; (6)~\citet{2005ApJ...621..793B}; (7)~\citet{2010ApJ...714..396A}; (8)~\citet{2011ApJ...729L..15T}; (9)~\citet{2000ApJ...545L..53H} + distance; (10)~\citet{1999AJ....118..926R}; (11)~\citet{2008ApJ...689..231V} + distance
}
\end{table*} 

\section{Results} \label{sect:results}

\subsection{Energy-loss model}

In this model we treat the magnetic field to be constant, $B(z)=B=\mbox{const.}$, implying no spatial dependence of the energy-loss term, $b(z)=1$. Using $d(z)=\alpha/b(z)$, as well as Eq. (\ref{eqn: substitution x}), the spatial coordinate $x(z)$ then scales as $$x(z(r))=\frac{z}{\alpha}=\frac{r_{\mbox{\tiny s}}-r}{\alpha} \, .$$

Using the parameters given in Table \ref{tab:parameters}, we calculate the X-ray intensity as a function of the projected radius according to Eq. (\ref{eqn: intensity distribution}). Here, the magnetic-field strength is a free parameter whose value can be chosen so that the filament widths match those found in the observations. In addition, we also determine the cut-off energy of the electron spectrum, which is connected to the magnetic field through Eq. (\ref{eqn:cutoff energy}). 

Reproducing the filament width for each SNR at a photon energy of 5~keV determines the magnetic-field strengths as given in Table \ref{tab:Constraints energy-loss model}. For our examples the downstream magnetic-field strength ranges from about $100~\mu$G up to about $500~\mu$G. Remnants with narrower filaments exhibit a higher downstream magnetic field. These magnetic fields then imply cut-off energies of the electron spectra in the energy range between 19~TeV and 37~TeV. The order of magnitude of the cut-off energies is in agreement with the results obtained from spectral modeling of the radio-to-X-ray spectra of young SNRs whose cut-off energies of their electron distribution must be in the TeV-band, since the cut-off frequencies are generally found in the X-ray band \citep{1999ApJ...525..368R}.  

\begin{figure}
  \resizebox{\hsize}{!}{\includegraphics[angle=0]{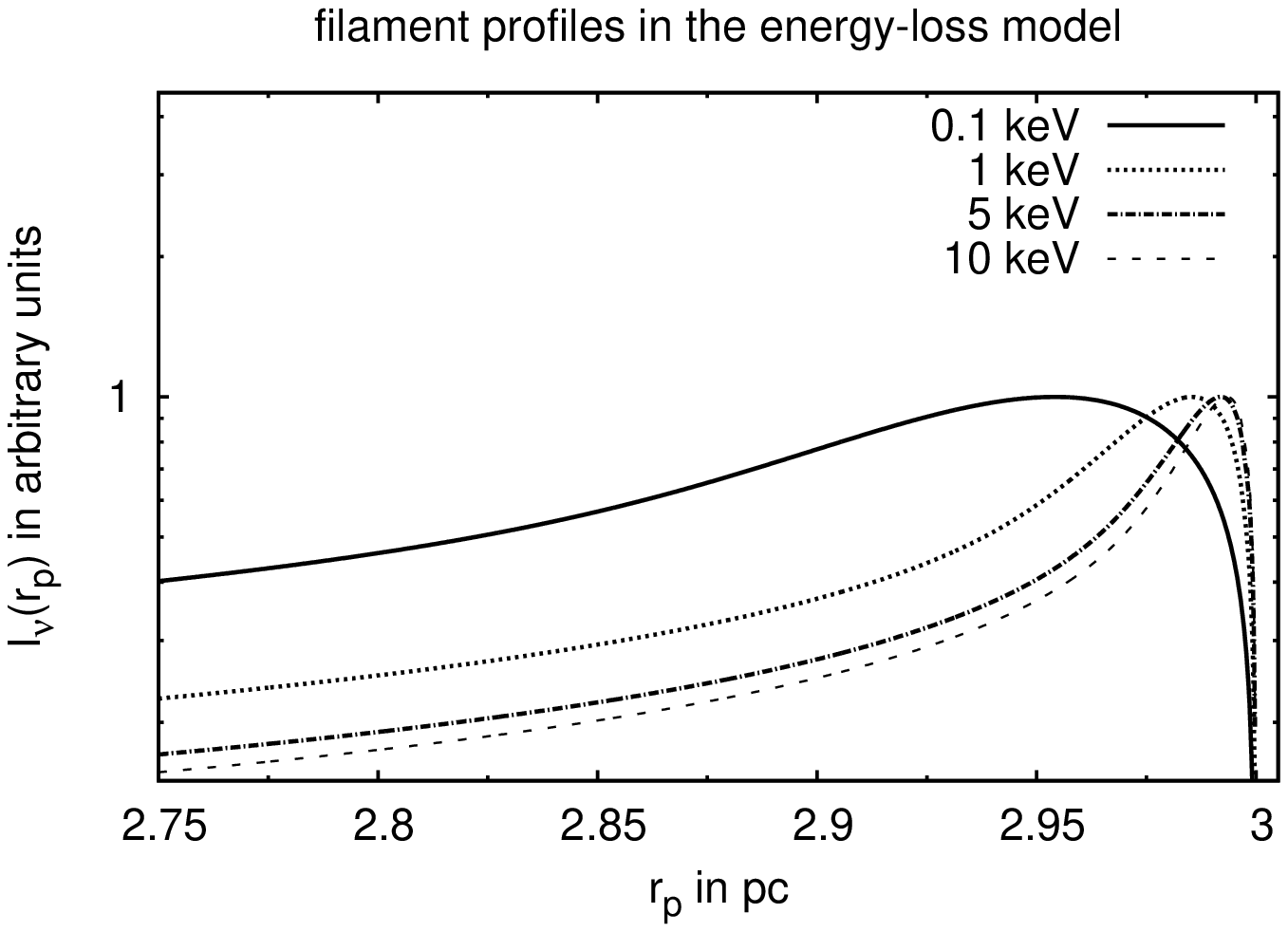}}
  \caption{Non-thermal X-ray intensity as a function of the projected radius calculated for four different X-ray energies with the parameters of Tycho given in Table \ref{tab:parameters}, as well as $B=310~\mu$G. The forward shock is located at $r_{\mbox{\tiny s}}=3~$pc.}
  \label{fig: filament profile loss}
\end{figure}

In Fig. (\ref{fig: filament profile loss}) we illustrate the profiles of the filaments for four different photon energies calculated with the parameters of Tycho. To be noted from the figure is a frequency dependence of the filament widths, which can be explained by the energy loss of the electrons in a constant magnetic field and by the advection process. The advection length represents the distance covered by the electrons within the synchrotron loss time, $\tau_{\mbox{\tiny syn}}=E/|\frac{\mbox{\tiny d}E}{\mbox{\tiny d}t}|$, and is given by
\begin{eqnarray}
l_{\mbox{\tiny ad}}&=&v\tau_{\mbox{\tiny syn}}=\frac{v_{\mbox{\tiny s}}}{4}\frac{9m^4c^7}{4e^4B^2E} \nonumber \\
&\simeq& (2\times 10^{-2}~\mbox{pc})\left(\frac{B}{200~\mu\mbox{G}}\right)^{-2} \left(\frac{v_{\mbox{\tiny s}}}{5000~\mbox{km}\,\mbox{s}^{-1}}\right) \left(\frac{E}{20~\mbox{TeV}}\right)^{-1}\, . \label{eqn:advection length}
\end{eqnarray}
Synchrotron radiation usually provides a continuum around a characteristic synchrotron frequency,
\begin{eqnarray} 
\nu_{\mbox{\tiny c}}&=&\frac{3\nu_{\mbox{\tiny L}}\gamma^2}{2}=\frac{3eBE^2}{4\pi m^3c^5} \nonumber \\
&\simeq & (1.3\times 10^{18}~\mbox{Hz}) \left(\frac{B}{200~\mu\mbox{G}}\right)\left(\frac{E}{20~\mbox{TeV}}\right)^2\;, \label{eqn: synchrotron frequency}
\end{eqnarray}
where $\nu_{\mbox{\tiny L}}$ and $\gamma $ are the Larmor frequency and the Lorentz factor of the accelerated electrons, respectively. Hence, the relation $E\propto \nu^{1/2}$. Therefore, a dependence of the width on the frequency of the radiation of the form $l_{\mbox{\tiny ad}}\propto \nu^{-1/2}$ would result. But the electrons are also affected by diffusion. With the Bohm diffusion coefficient, $D=r_{\mbox{\tiny L}}c/3$, one thus obtains for the corresponding diffusion length
\begin{equation} \label{eqn:diffusion length}
l_{\mbox{\tiny diff}}=\sqrt{D\tau_{\mbox{\tiny syn}}}=\sqrt{\frac{3m^4c^8}{4e^5B^3}}\simeq (1.3\times 10^{-2}~\mbox{pc}) \left(\frac{B}{200~\mu\mbox{G}}\right)^{-3/2}\, ,
\end{equation}
which does not depend on the electron energy. Equating the advection and diffusion length, the relation
\begin{equation} \label{eqn: critical energy}
E_{\mbox{\tiny c}}\simeq (31~\mbox{TeV}) \left(\frac{v_{\mbox{\tiny s}}}{5000~\mbox{km}\,\mbox{s}^{-1}}\right)\left(\frac{B}{200~\mu\mbox{G}}\right)^{-1/2}\, 
\end{equation}
can be derived. Electrons with energies $E>E_{\mbox{\tiny c}}$ can stream farther from the shock than advection alone would allow. According to Eq. (\ref{eqn: synchrotron frequency}) and Eq.(\ref{eqn: critical energy}), the characteristic synchrotron frequency for the electrons with $E>E_{\mbox{\tiny c}}$ is then higher than
\begin{equation}
\nu_{\mbox{\tiny c}}(E_{\mbox{\tiny c}})\simeq (3.1\times 10^{18}~\mbox{Hz})\left(\frac{v_{\mbox{\tiny s}}}{5000~\mbox{km}\,\mbox{s}^{-1}}\right)^2\;,
\end{equation}
corresponding to a X-ray energy of about 13~keV, if $v_{\mbox{\tiny s}}=5000~$km/s. Hence, the filament profiles at higher photon energies, which need the most energetic electrons, show approximately the same behaviour as can be seen in Fig. (\ref{fig: filament profile loss}) for the 5-keV and 10-keV profile. 

In addition, the advection and diffusion length also explain the relation between the filament widths and the corresponding magnetic fields given in Table \ref{tab:Constraints energy-loss model}. Narrower filaments require a faster decrease in the synchrotron emissivity, implying a shorter advection and diffusion length. And according to Eq. (\ref{eqn:advection length}) and Eq. (\ref{eqn:diffusion length}), this is given for higher magnetic-field strengths.

\begin{figure}
  \resizebox{\hsize}{!}{\includegraphics[angle=0]{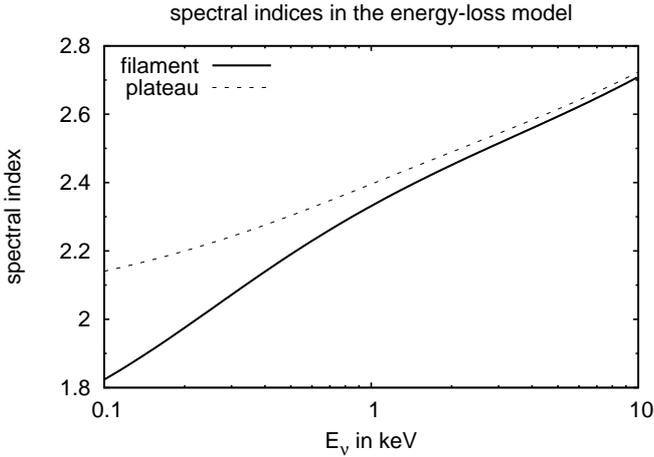}}
  \caption{Photon spectral indices of the spectra of filament and plateau using the parameters of Tycho.}
  \label{fig: indices loss}
\end{figure}

Now, equipped with the X-ray intensity distribution establishing the filament profiles, and the volume emissivity, we calculate the spectra of the filament and plateau for each of our examples. To obtain the filament spectrum, we integrate the intensity along the projected radius from $r_{\mbox{\tiny p}}=r_{\mbox{\tiny s}}$ up to $r_{\mbox{\tiny p}}=r_{\mbox{\tiny s}}-w$. Because we do not know the electron source strength, $q_0$, we are not interested in absolute fluxes. However, we can show the qualitative behaviour described by the appropriate photon spectral indices, which should be sufficient for the comparison. Assuming the photon spectra to show a power-law characteristic, $N_\nu=F_\nu/h\nu\propto \nu^{-\Gamma}$, where $F_\nu$ and $\Gamma$ are the flux and photon spectral index, respectively, we can then describe the spectra between the photon energies $h\nu$ and $h\nu'$ through
\begin{equation} \label{eqn:spectral index}
\Gamma=\frac{\ln\left(\frac{\nu F_{\nu'}}{\nu' F_\nu}\right)}{\ln\left(\frac{\nu}{\nu'}\right)} \; .
\end{equation}
Here, we also want to calculate the differences between the photon spectral index of the filament spectrum, $\Gamma_{\mbox{\tiny f}}$, and that of the plateau emission, $\Gamma_{\mbox{\tiny p}}$, which are given at three different photon energies, $E_\nu$, in Table \ref{tab:Constraints energy-loss model}. Additionally, in Fig. (\ref{fig: indices loss}) we show the photon spectral indices calculated with the parameters of Tycho. 

As can be seen from Table \ref{tab:Constraints energy-loss model}, as well as from Fig. (\ref{fig: indices loss}), the spectra at higher photon energies rarely differ significantly. The plateau shows a steeper spectrum at lower photon energies. Up to a photon energy of 1~keV the difference between the indices of filament and plateau is in the range $0.05-0.34$, whereas at energies higher than 1~keV the difference is always smaller than 0.1. This property can be explained by the effective radiation of the energetic electrons in the enhanced magnetic field. Accordingly, the most energetic electrons lose all of their energy inside the filaments, implying that the regions farther from the shock provide almost no contributions to the total emission of hard X-rays, so that both filament and plateau show nearly the same behaviour.

In Table \ref{tab:Constraints energy-loss model} we also show the indices of the filaments at three different photon energies. One can see that the parameters of Cas A, Tycho, Kepler and SN 1006 lead to the same spectral behaviour, if their shock velocities are similar. However, using the shock velocities measured in the northwestern limb of SN 1006 and resulting from the upper limit to the distance to Kepler, it turns out that the filament spectrum is softer and harder, respectively, than in the former case.

\begin{table*}
\caption{Constraints on the downstream magnetic-field strength, cut-off energy of the electron spectrum, photon spectral index of the filament, and the difference between the photon indices of the spectra of filament and plateau at three different photon energies, $E_\nu$, for four young SNRs. These values are calculated using the energy-loss model and the parameters given in Table \ref{tab:parameters}.} \label{tab:Constraints energy-loss model}
\centering
\begin{tabular}{c c c c c c c c c}
\hline \hline
SNR       & $B$      & $E_{\mbox{\tiny cut}}$ & \multicolumn{3}{c}{$\Gamma_{\mbox{\tiny f}}$} & \multicolumn{3}{c}{$\Gamma_{\mbox{\tiny p}}-\Gamma_{\mbox{\tiny f}}$} \\ \cmidrule(lr){4-6} \cmidrule(lr){7-9}
          & [$\mu$G] & [TeV] & $E_\nu=0.1~$keV & $E_\nu=1~$keV & $E_\nu=10~$keV & $E_\nu=0.1~$keV & $E_\nu=1~$keV & $E_\nu=10~$keV \\
\hline
SN 1006\tablefootmark{a} & 130   & 36   & 1.81   & 2.33   & 2.71   & 0.31   & 0.07   & 0.02  \\
                         & (110) & (23) & (1.99) & (2.49) & (2.94) & (0.25) & (0.05) & (0.01) \\
Cas A                    & 520   & 19   & 1.83   & 2.33   & 2.70   & 0.31   & 0.06   & 0.01   \\
Tycho                    & 310   & 24   & 1.82   & 2.33   & 2.71   & 0.32   & 0.06   & 0.01    \\
Kepler\tablefootmark{b}  & 250   & 26   & 1.81   & 2.32   & 2.70   & 0.31   & 0.07   & 0.01     \\
                         & (230) & (37) & (1.75) & (2.23) & (2.61) & (0.34) & (0.08) & (0.02)    \\
\hline
\end{tabular}
\tablefoot{
\tablefoottext{a}{The values in brackets were calculated using the shock velocity $v_{\mbox{\tiny s}}=2900~$km/s as measured in the northwestern limb.}
\tablefoottext{b}{The values in brackets were calculated using the upper limit of 6.4~kpc to the distance.}
}
\end{table*}

\subsection{Magnetic-field damping model}

As already mentioned above, in the model of magnetic-field damping we assume the magnetic-field strength to follow a profile described in Eq. (\ref{eqn:magnetic field}), which can also be written as $$B(z)=B_{\mbox{\tiny min}}\left[1+\frac{B_{\mbox{\tiny max}}-B_{\mbox{\tiny min}}}{B_{\mbox{\tiny min}}}\exp\left(-\frac{z}{l_{\mbox{\tiny d}}}\right)\right] \,.$$ The spatial dependence of the energy-loss term then obeys the relation $$b(z)=\left[1+\frac{B_{\mbox{\tiny max}}-B_{\mbox{\tiny min}}}{B_{\mbox{\tiny min}}}\exp\left(-\frac{z}{l_{\mbox{\tiny d}}}\right)\right]^2 \, ,$$ corresponding to a spatial variation of the diffusion coefficient $$d(z)=\alpha \left[1+\frac{B_{\mbox{\tiny max}}-B_{\mbox{\tiny min}}}{B_{\mbox{\tiny min}}}\exp\left(-\frac{z}{l_{\mbox{\tiny d}}}\right)\right]^{-2} \,,$$ because the product $b(z)d(z)$ must be constant, as required by Eq. (\ref{eqn:alpha}). According to Eq. (\ref{eqn: substitution x}), the spatial coordinate $x(z)$ then scales as 
\begin{eqnarray}
x(z)&=&\frac{1}{\alpha}\left\{z+\frac{l_{\mbox{\tiny d}}}{2}\frac{(B_{\mbox{\tiny max}}-B_{\mbox{\tiny min}})^2}{B_{\mbox{\tiny min}}^2}\left[1-\exp\left(-\frac{2z}{l_{\mbox{\tiny d}}}\right)\right]\right.  \nonumber \\
    && \hspace{2cm}\left. +2l_{\mbox{\tiny d}}\frac{B_{\mbox{\tiny max}}-B_{\mbox{\tiny min}}}{B_{\mbox{\tiny min}}}\left[1-\exp\left(-\frac{z}{l_{\mbox{\tiny d}}}\right)\right] \right\}\,.
\end{eqnarray}

Again, using the parameters given in Table \ref{tab:parameters}, we calculate the X-ray intensity as a function of the projected radius. In this case, the damping length and the maximum field strength, $B_{\mbox{\tiny max}}$, are free parameters that can be chosen so that the filament widths match those observed. According to the calculation of \citet{2005ApJ...626L.101P}, the damping length should be in the range $l_{\mbox{\tiny d}}=10^{16}-10^{17}~$cm ($l_{\mbox{\tiny d}}=0.003-0.03~$pc). To also investigate the influence of the damping length on the results, we perform the calculation using two different values of $l_{\mbox{\tiny d}}$ in each of our examples. Here, the larger value used for $l_{\mbox{\tiny d}}$ may describe weak magnetic-field damping, whereas the smaller one may cause a strong damping of the field. But note that energy losses are still included.  

Reproducing the filament width of each SNR at a photon energy of 5~keV requires the maximum field strengths given in Table \ref{tab:Constraints magnetic damping model}. Depending on the damping length, the maximum field strength can be found in the range between 50~$\mu$G and 260~$\mu$G, implying, according to Eq. (\ref{eqn:cutoff energy}), cut-off energies between 27~TeV and 62~TeV. Furthermore, it turns out that an increased damping length requires an increased field strength, because the electrons radiate efficiently in a larger volume, resulting in wider filaments. To retain the observed filament widths, it is then necessary to have a higher magnetic field that, on the other hand, also leads to a smaller cut-off energy of the electron distribution.

If the damping length is too small, the observed filament widths cannot be realized for any maximum field strength. In these cases the intensity first decreases but then increases again even for the 5-keV profile, so that the typical shape of the filament profiles is not given anymore. Therefore, we use damping lengths in the case of strong damping, for which the profiles just still exhibit the typical shape. For instance, using the given parameters of SN 1006, the damping length used in the calculation should not be smaller than 0.02~pc.

\begin{figure}
  \resizebox{\hsize}{!}{\includegraphics[angle=0]{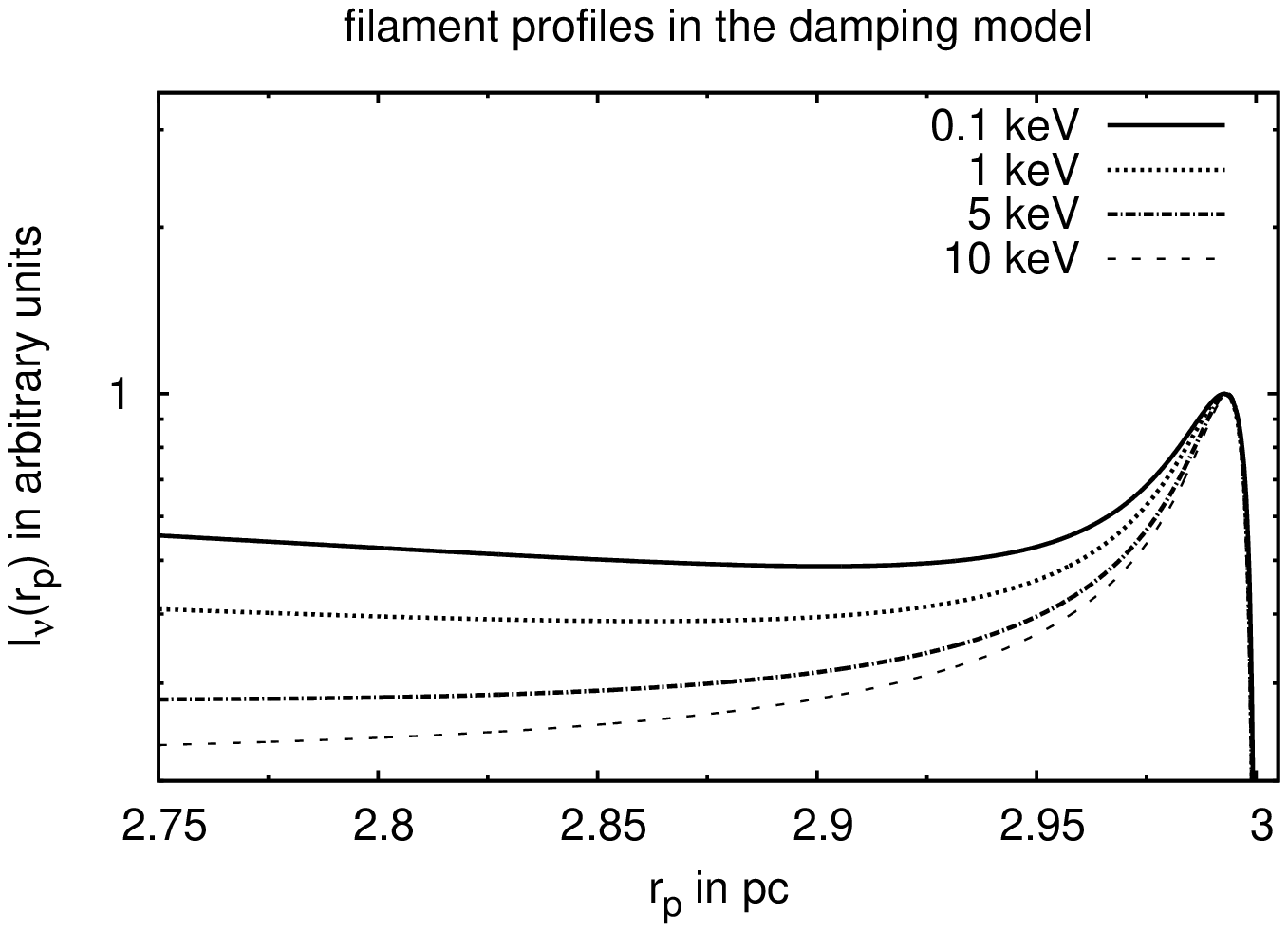}}
  \caption{Non-thermal X-ray intensity as a function of the projected radius calculated for four different X-ray energies with the parameters of Tycho given in Table \ref{tab:parameters}, as well as $l_{\mbox{\tiny d}}=0.008~$pc and $B_{\mbox{\tiny max}}=85~\mu$G. The forward shock is located at $r_{\mbox{\tiny s}}=3~$pc.}
  \label{fig: filament profile damping}
\end{figure}

The filament profiles calculated with the parameters of Tycho for four different photon energies are illustrated in Fig. (\ref{fig: filament profile damping}). As can be seen from the figure, there is also a dependence of the filaments on the frequency of the X-rays. This dependence is based on the spatial variation of the magnetic-field strength. Only in regions very close to the shock front the electrons radiate in fields of high magnitude, so that even the most energetic of them can emit photons of several keV in energy only in a small volume. The electrons remain energetic when they propagate into the downstream region, where they radiate at lower frequencies. Hence, we expect increased emission of low-energy X-rays in regions farther from the shock. For instance, using the cut-off energy of the electron spectrum in Tycho obtained for the case of strong damping, $E_{\mbox{\tiny cut}}=45~$TeV, one finds that according to Eq. (\ref{eqn: synchrotron frequency}), even the most energetic electrons located in a magnetic field of $B=10~\mu$G have their synchrotron continuum around the characteristic frequency $\nu_{\mbox{\tiny c}}(E_{\mbox{\tiny cut}})=3.3\times 10^{17}~$Hz, corresponding to about 1.4~keV in X-ray energy. This issue can be seen for the 0.1-keV and 1-keV profile in Fig. (\ref{fig: filament profile damping}). The X-ray intensity does not decrease with decreasing projected radius as happens in the energy-loss model, but remains nearly constant and even increases, respectively.

\begin{figure}
  \resizebox{\hsize}{!}{\includegraphics[angle=0]{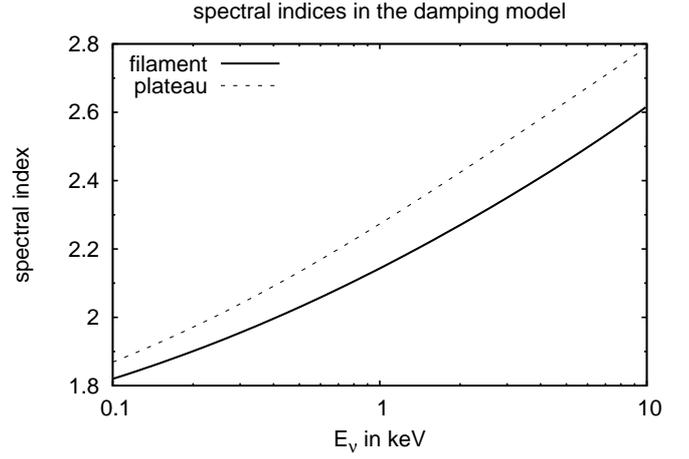}}
  \caption{Photon spectral indices of the spectra of filament and plateau using the parameters of Tycho.}
  \label{fig: indices damping}
\end{figure}

Using the X-ray intensity distribution, as well as the volume emissivity, we now calculate the spectra of filament and plateau for each example. Again, we integrate the intensity along the projected radius between $r_{\mbox{\tiny p}}=r_{\mbox{\tiny s}}$ and $r_{\mbox{\tiny p}}=r_{\mbox{\tiny s}}-w$ in order to obtain the filament spectrum. The difference between the photon spectral indices of filament and plateau, $\Gamma_{\mbox{\tiny f}}-\Gamma_{\mbox{\tiny p}}$, at three photon energies is given in Table \ref{tab:Constraints magnetic damping model}, whereas in Fig. (\ref{fig: indices damping}) we illustrate the photon spectral indices calculated with the parameters of Tycho.

As can be seen again, the spectrum of the plateau is steeper than that of the filament. However, the difference between the spectra of filament and plateau depends on the chosen damping length. At relatively large damping lengths (weak damping) the difference between the indices over the hole energy range is smaller than 0.1, whereas at smaller damping lengths (strong damping) it also takes values in the range 0.1-0.2. The small differences at larger damping lengths are due to the higher magnetic fields that need to be chosen in order to retain the observed filament widths. Hence, considerable energy losses have to be taken into account. Similarly to the energy-loss model, this results in an almost equal behaviour of the spectra of filament and plateau at high photon energies. In contrast, the magnetic fields used at smaller damping lengths are low enough to result in spectra that show significant differences among each other.

Finally, one can also see from Table \ref{tab:Constraints magnetic damping model} that the filament spectrum becomes steeper with decreasing damping length, in particular at small X-ray energies. This is due to the lower magnetic-field strengths used in that case. Although the weaker magnetic fields imply higher cut-off energies, which harden the spectra, their influence is not sufficient enough to result in spectra similar to those found at larger damping lengths. Besides, as in the energy-loss model and independently of the damping length, Cas A, Tycho, Kepler and SN 1006 show roughly the same spectral behaviour, if the shock velocities are similar, whereas the filament spectrum obtained from the shock velocity of the northwestern limb of SN 1006 has a steeper profile. In contrast, the upper limit to the distance to Kepler implies a harder filament spectrum.

\begin{table*}
\caption{Constraints on the maximum magnetic-field strength, cut-off energy of the electron spectrum, photon spectral index of the filament, and the difference between the photon indices of the spectra of filament and plateau at three different photon energies, $E_\nu$, for four young SNRs. These values are calculated using the magnetic-field damping model and the parameters given in Table \ref{tab:parameters}.} \label{tab:Constraints magnetic damping model}
\centering
\begin{tabular}{c c c c c c c c c c}
\hline \hline
\multicolumn{10}{c}{Weak Damping} \\
\hline
SNR & $l_{\mbox{\tiny d}}$ & $B_{\mbox{\tiny max}}$ & $E_{\mbox{\tiny cut}}$ & \multicolumn{3}{c}{$\Gamma_{\mbox{\tiny f}}$} & \multicolumn{3}{c}{$\Gamma_{\mbox{\tiny f}}-\Gamma_{\mbox{\tiny p}}$} \\ \cmidrule(lr){5-7} \cmidrule(lr){8-10}
    & [pc] & [$\mu$G] & [TeV] & $E_\nu=0.1~$keV & $E_\nu=1~$keV & $E_\nu=10~$keV & $E_\nu=0.1~$keV & $E_\nu=1~$keV & $E_\nu=10~$keV \\
    \hline
SN 1006 & 0.03   & 65    & 50   & 1.78   & 2.12   & 2.64   & 0.04   & 0.08   & 0.07  \\
        & (0.03) & (57)  & (32) & (1.93) & (2.35) & (2.95) & (0.05) & (0.09) & (0.08) \\
Cas A   & 0.015  & 260   & 27   & 1.71   & 1.98   & 2.51   & 0.04   & 0.02   & 0.01   \\
Tycho   & 0.02   & 150   & 34   & 1.71   & 2.00   & 2.57   & 0.06   & 0.04   & 0.02   \\
Kepler  & 0.03   & 135   & 36   & 1.71   & 2.01   & 2.59   & 0.05   & 0.04   & 0.02   \\
        & (0.03) & (115) & (52) & (1.67) & (1.93) & (2.43) & (0.04) & (0.05) & (0.02) \\
\hline
\multicolumn{10}{c}{Strong Damping} \\
\hline
SNR & $l_{\mbox{\tiny d}}$ & $B_{\mbox{\tiny max}}$ & $E_{\mbox{\tiny cut}}$ & \multicolumn{3}{c}{$\Gamma_{\mbox{\tiny f}}$} & \multicolumn{3}{c}{$\Gamma_{\mbox{\tiny f}}-\Gamma_{\mbox{\tiny p}}$} \\ \cmidrule(lr){5-7} \cmidrule(lr){8-10}
    & [pc] & [$\mu$G] & [TeV] & $E_\nu=0.1~$keV & $E_\nu=1~$keV & $E_\nu=10~$keV & $E_\nu=0.1~$keV & $E_\nu=1~$keV & $E_\nu=10~$keV \\ 
    \hline
SN 1006\tablefootmark{a} & 0.02    & 64   & 51   & 1.81   & 2.16   & 2.66   & 0.04   & 0.10   & 0.13   \\
                         & (0.02)  & (56) & (32) & (1.97) & (2.39) & (2.96) & (0.04) & (0.11) & (0.13) \\
Cas A                    & 0.004   & 115  & 40   & 1.81   & 2.12   & 2.58   & 0.07   & 0.16   & 0.19   \\
Tycho                    & 0.008   & 85   & 45   & 1.82   & 2.14   & 2.61   & 0.05   & 0.13   & 0.17   \\
Kepler\tablefootmark{b}  & 0.01    & 80   & 47   & 1.81   & 2.14   & 2.62   & 0.04   & 0.11   & 0.14    \\
                         & (0.012) & (80) & (62) & (1.74) & (2.03) & (2.48) & (0.03) & (0.09) & (0.12)   \\
\hline
\end{tabular}
\tablefoot{
\tablefoottext{a}{The values in brackets were calculated using the shock velocity $v_{\mbox{\tiny s}}=2900~$km/s as measured in the northwestern limb.}
\tablefoottext{b}{The values in brackets were calculated using the upper limit of 6.4~kpc to the distance.}
}
\end{table*}

\section{Conclusions}

Compared to the magnetic-field damping model, the spectra of filament and plateau obtained in the energy-loss model exhibit larger spectral indices. This can be explained by the considerable energy losses leading to the evolution towards a softer electron distribution in the energy-loss model, and hence, resulting in X-ray spectra that are softer than those obtained in the damping model.   

In case of a weak magnetic-field damping the difference between the spectral indices of filament and plateau over the full X-ray spectrum is smaller than 0.1, which is probably to small to be detectable. Only if there is a strong damping, our calculation suggests a measurable difference between the spectra in some SNRs, since the difference between the spectral indices of filament and plateau can take values of almost 0.2 at X-ray energies higher than 1~keV.

In the energy-loss model the difference between the indices of filament and plateau above the X-ray energy of 1~keV is even smaller than 0.1, so that a possible detection can be excluded here, too. On the other hand, the difference between the indices below 1~keV is larger than 0.1, and at a photon energy of 0.1~keV it is even approximately 0.3. This might suggest that there is a measurable difference in the spectra of filament and plateau at small X-ray energies, if the filaments are limited by energy losses. However, on account of the interstellar photoelectric absorption of the soft X-rays, these different spectral characteristics are probably not detectable, too. Furthermore, the plasma downstream of the forward shock is at high temperature, implying also thermal emission contributing to the soft X-ray band, and thus, complicating a clear identification of the non-thermal emission.

Hence, if there is no measurable difference between the spectra of filament and plateau, it is not possible to make definite predictions from the comparison of the spectra whether the filaments are limited by energy losses of the radiating electrons or by damping of the magnetic field. But if a significant difference appears, our calculations then suggest that the filaments are limited by the magnetic field itself.

It should be noted that our results presented here have been derived using Bohm diffusion. According to Eq. (\ref{eqn:diffusion length}), a larger diffusion coefficient with gyrofactor $\eta>1$ would imply a larger diffusion length, resulting in significant widening of the filaments, because now, the regions farther from the shock contain a sufficient number of high-energy electrons contributing to the intensity. Widening must then be compensated by a higher magnetic field to retain the observed filament widths. Hence, the magnetic-field strengths derived in our models represent lower limits for the chosen parameters. The calculation then shows that a larger diffusion coefficient results in softer spectra due to a lower cut-off energy, which decreases with increasing gyrofactor. However, the final results regarding the differences in spectral indices do not change fundamentally.  

In a last step we want to compare the predictions derived here with observations. At first, we notice that, independently of the model, the parameters from the remnants of Cas A, Tycho, Kepler and SN 1006 lead to nearly the same spectral behaviour in case of similar shock velocities, as can be seen from the spectral indices in Table \ref{tab:Constraints energy-loss model} and Table \ref{tab:Constraints magnetic damping model}. However, the analysis of the filament spectra of these remnants reported by \citet{2003ApJ...589..827B,2005ApJ...621..793B} reveals significant differences among the spectral indices obtained from the fit of an absorbed power-law model. Compared to our spectra whose calculation has been done using an injection index resulting from an unmodified shock ($s=2$), the observation may be an indication for different electron injection indices in these remnants, implying shocks that are differently affected by non-linear effects due to differences in efficiency in the particle acceleration.

Regarding the magnetic-field strengths, we take, as an example for comparison with our results, the non-thermal filaments of Cas A analysed by \citet{2010ApJ...714..396A}. From the best-fit parameters used to fit the observed filament spectra, the magnetic field has been derived to be in the range $(30-70)~\mu$G. These values are consistent with those derived from the magnetic-field damping model, in which the magnetic field varies, according to Eq. (\ref{eqn:magnetic field}) and the values from Table \ref{tab:Constraints magnetic damping model}, between the field strengths $(10-260)~\mu$G for weak damping and $(10-115)~\mu$G for strong damping, respectively. For comparison, the constant magnetic field derived from the energy-loss model is several times higher, $B=520~\mu$G. This might suggest that the non-thermal filaments of Cas A are limited by the damping of the magnetic field.

Another comparison concerns the magnetic fields in SN 1006 and Tycho. Using the data from radio up to TeV-observations, \citet{2010A&A...516A..62A} have analysed the multi-wavelength spectrum of SN 1006 in the framework of a leptonic and hadronic origin for the gamma-ray emission, giving a magnetic-field of $\sim 30~\mu$G in the leptonic model and a magnetic field of $\sim 120~\mu$G in the hadronic model, respectively. Moreover, combining radio and X-ray data with recent TeV-observations performed with the VERITAS instrument, the magnetic field of Tycho has been estimated to be $\sim 80~\mu$G in a leptonic-dominated model, whereas a hadronic dominated model yields a magnetic field of $\sim 230~\mu$G \citep{2011ApJ...730L..20A}. Compared to our model predictions given in Table \ref{tab:Constraints energy-loss model} and Table \ref{tab:Constraints magnetic damping model}, we notice that the magnetic fields derived from the energy-loss model are in good agreement with those estimated from the hadronic model used to describe the observed spectra of SN 1006 and Tycho. In contrast, the predictions from the magnetic-field damping model suggest the leptonic model for the origin of the gamma-ray emission from these remnants. It should be noted that current gamma-ray observations do not reach the spatial resolution of those done in X-rays, so that the magnetic fields estimated using gamma-ray observations of SN 1006 and Tycho are averages over a region much larger than the filaments, implying that the observed values do not necessarily match those found for the filaments.

To discriminate between the energy-loss model and magnetic-field damping model, and hence between a leptonic and a hadronic origin of TeV-band gamma-ray emission, one may either search for differences between X-ray spectra of filaments and plateau, as calculated in this paper, or perform gamma-ray observations with higher spatial resolution.

\subsection*{Acknowledgement}

We acknowledge support by the "Helmholtz Alliance for Astroparticle Phyics HAP" funded by the Initiative and Networking Fund of the Helmholtz Association.    

\bibliographystyle{aa}
\bibliography{References}

\end{document}